\documentstyle[amssymb,floats,aps,prb,epsf]{revtex}
\epsfclipon
\begin{document}
\twocolumn[\hsize\textwidth\columnwidth\hsize\csname
@twocolumnfalse\endcsname

\draft

\title{Magnetic nature of superconductivity in doped cuprates}
\author{Shiping Feng, Tianxing Ma, and Huaiming Guo}
\address{Department of Physics, Beijing Normal University, Beijing
100875, China}
\maketitle
%\date{\today}
\begin{abstract}
Within the kinetic energy driven superconducting mechanism, the
magnetic nature of cuprate superconductors is discussed. It is
shown that the superconducting state is controlled by both charge
carrier gap function and quasiparticle coherent weight. This
quasiparticle coherent weight grows linearly with the hole doping
concentration in the underdoped and optimally doped regimes, and
then decreases with doping in the overdoped regime, which leads to
that the maximal superconducting transition temperature occurs
around the optimal doping, and then decreases in both underdoped
and overdoped regimes. Within this framework, we calculate the
dynamical spin structure factor of cuprate superconductors, and
reproduce all main features of inelastic neutron scattering
experiments, including the energy dependence of the incommensurate
magnetic scattering at both low and high energies and commensurate
resonance at intermediate energy.
\end{abstract}

]

\bigskip

\narrowtext

\section{Introduction}

The interplay between the strong electron correlation and
superconductivity is one of the most important problems raised by
the discovery of cuprate superconductors \cite{anderson2}. After
intensive investigations over more than a decade, it has become
clear that the strong electron correlation in doped cuprates plays
a crucial role not only for the unusual normal-state behavior but
also for the superconducting (SC) mechanism
\cite{anderson2,anderson3,laughlin}. The parent compound of
cuprates superconductors is a Mott insulator with the
antiferromagnetic (AF) long-range order (AFLRO), then changing the
carrier concentration by ionic substitution or increasing the
oxygen content turns these compounds into the SC-state leaving the
AF short-range correlation (AFSRC) still intact \cite{kastner}. As
a function of the hole doping concentration, the SC transition
temperature reaches a maximum in the optimal doping, and then
decreases in both underdoped and overdoped regimes \cite{tallon}.
Moreover, this SC transition temperature is dependence of both
charge carrier gap parameter and quasiparticle coherent weight
\cite{ding}, which strongly suggests that the quasiparticle
coherence plays an important role in superconductivity.

By virtue of systematic studies using the nuclear magnetic
resonance, and muon spin rotation techniques, particularly the
inelastic neutron scattering, the doping and energy dependent
magnetic excitations in doped cuprates in the SC-state have been
well established: (a) at low energy, the incommensurate (IC)
magnetic scattering peaks are shifted from the AF wave vector
[$\pi$,$\pi$] to four points [$(1\pm\delta)\pi, \pi$] and
[$\pi,(1\pm\delta)\pi$] (in units of inverse lattice constant)
with $\delta $ as the incommensurability parameter
\cite{dai,yamada,wakimoto}; (b) then with increasing energy these
IC magnetic scattering peaks are converged on the commensurate
[$\pi $,$\pi $] resonance peak at intermediate energy
\cite{dai,bourges0,bourges,arai}; and (c) well above this
resonance energy, the continuum of magnetic excitations peaked at
IC positions in the diagonal direction are oberved
\cite{hayden,stock,tranquada}. It has been emphasized that the
geometry of these IC magnetic excitations is two-dimensional
\cite{hinkov,hayden}. Although some of these magnetic properties
have been observed in the normal-state, these IC magnetic
scattering and commensurate resonance are the main new feature
that appears into the SC-state. Moreover, AFSRC coexists with the
SC-state in the whole SC regime \cite{wakimoto}, and the unusual
magnetic excitations at high energy have energies greater than the
SC pairing energy, are present at the SC transition temperature,
and have spectral weight far exceeding that of the resonance
\cite{hayden,stock}. These provide a clear link between the charge
carrier pairing mechanism and magnetic excitations in cuprate
superconductors.

Recently, we \cite{feng1} have discussed the kinetic energy driven
SC mechanism in doped cuprates based on the charge-spin separation
(CSS) fermion-spin theory \cite{feng2}, where the dressed holons
interact occurring directly through the kinetic energy by
exchanging dressed spin excitations, leading to a net attractive
force between dressed holons, then the electron Cooper pairs
originating from the dressed holon pairing state are due to the
charge-spin recombination, and their condensation reveals the SC
ground-state. The SC transition temperature is proportional to the
hole doping concentration in the underdoped regime. However, an
obvious weakness is that the SC transition temperature is too
high, and not suppressed in the overdoped regime \cite{feng1}. In
this paper, we study the magnetic nature of the kinetic energy
superconductivity in doped cuprates along with this line. A short
version of this work was published earlier \cite{feng0}. One of
our main results is that the SC transition temperature is
suppressed to low temperatures by considering the quasiparticle
coherence, and therefore the SC transition temperature is
controlled by both charge carrier gap function and quasiparticle
coherent weight. This quasiparticle coherent weight is closely
related to the dressed holon self-energy from the dressed spin
pair bubble, and grows linearly with increasing doping in the
underdoped and optimally doped regimes, then decreases with
increasing doping in the overdoped regime, which leads to that the
maximal SC transition temperature occurs around the optimal
doping, and then decreases in both underdoped and overdoped
regimes. Within this SC mechanism, we give a theoretical
explanation of inelastic neutron scattering experiments on cuprate
superconductors
\cite{dai,bourges0,bourges,arai,hayden,stock,tranquada} in terms
of the collective mode in the dressed holon particle-particle
channel.

The paper is organized as follows. The interplay between the
quasiparticle coherence and superconductivity is discussed in Sec.
II. In Sec. III, we calculate explicitly the dynamical spin
structure factor of cuprate superconductors, and reproduce all
main features found in experiments in the SC-state
\cite{dai,yamada,wakimoto,bourges0,bourges,arai,hayden,stock,tranquada},
including the energy dependence of the IC magnetic scattering at
both low and high energies and commensurate $[\pi,\pi ]$ resonance
at intermediate energy. Sec. IV is devoted to a summary and
discussions.

\section{Interplay between the quasiparticle coherence and
superconductivity}

In doped cuprates, the single common feature is the presence of
the two-dimensional CuO$_{2}$ plane \cite{kastner}, it is believed
that the relatively high SC transition temperature is closely
related to doped CuO$_{2}$ planes. It has been argued that the
essential physics of the doped CuO$_{2}$ plane is contained in the
$t$-$J$ model on a square lattice \cite{anderson2},
\begin{equation}
H=-t\sum_{i\hat{\eta}\sigma}C^{\dagger}_{i\sigma}
C_{i+\hat{\eta}\sigma}+\mu \sum_{i\sigma}
C^{\dagger}_{i\sigma}C_{i\sigma}+J\sum_{i\hat{\eta}}{\bf S}_{i}
\cdot {\bf S}_{i+\hat{\eta}},
\end{equation}
with $\hat{\eta}=\pm\hat{x},\pm\hat{y}$, $C^{\dagger}_{i\sigma}$
($C_{i\sigma}$) is the electron creation (annihilation) operator,
${\bf S}_{i}=C^{\dagger}_{i}{\vec\sigma} C_{i}/2$ is spin operator
with ${\vec\sigma}=(\sigma_{x}, \sigma_{y},\sigma_{z})$ as Pauli
matrices, and $\mu$ is the chemical potential. The $t$-$J$ model
(1) is subject to an important local constraint to avoid the
double occupancy, i.e., $\sum_{\sigma}C^{\dagger}_{i\sigma}
C_{i\sigma} \leq 1$. In the $t$-$J$ model, the strong electron
correlation manifests itself by this single occupancy local
constraint, and therefore the crucial requirement is to impose
this local constraint. This local constraint can be treated
properly in analytical calculations within the CSS fermion-spin
theory \cite{feng2}, where the constrained electron operators are
decoupled as, $C_{i\uparrow}= h^{\dagger}_{i\uparrow} S^{-}_{i}$
and $C_{i\downarrow}=h^{\dagger}_{i\downarrow}S^{+}_{i}$, with the
spinful fermion operator $h_{i\sigma}= e^{-i\Phi_{i\sigma}}h_{i}$
describes the charge degree of freedom together with some effects
of the spin configuration rearrangements due to the presence of
the hole itself (dressed holon), while the spin operator $S_{i}$
describes the spin degree of freedom (dressed spin), then the
electron local constraint for the single occupancy is satisfied in
analytical calculations \cite{feng2}. In this CSS fermion-spin
representation, the low-energy behavior of the $t$-$J$ model (1)
can be expressed as \cite{feng1,feng2,feng0},
\begin{eqnarray}
H&=&-t\sum_{i\hat{\eta}}(h_{i\uparrow}S^{+}_{i}
h^{\dagger}_{i+\hat{\eta}\uparrow}S^{-}_{i+\hat{\eta}}+
h_{i\downarrow}S^{-}_{i}h^{\dagger}_{i+\hat{\eta}\downarrow}
S^{+}_{i+\hat{\eta}}) \nonumber\\
&-&\mu\sum_{i\sigma}h^{\dagger}_{i\sigma} h_{i\sigma}+J_{{\rm
eff}}\sum_{i\hat{\eta}}{\bf S}_{i}\cdot {\bf S}_{i+\hat{\eta}},
\end{eqnarray}
with $J_{{\rm eff}}=(1-x)^{2}J$, and $x=\langle
h^{\dagger}_{i\sigma}h_{i\sigma}\rangle=\langle h^{\dagger}_{i}
h_{i}\rangle$ is the hole doping concentration. As a consequence,
the kinetic energy (t) term in the $t$-$J$ model has been
expressed as the dressed holon-spin interaction, which reflects
that even kinetic energy term in the $t$-$J$ model has strong
Coulombic contributions due to the restriction of single occupancy
of a given site. This dressed holon-spin interaction is quite
strong, and we \cite{feng1,feng0} have shown in terms of
Eliashberg's strong coupling theory \cite{eliashberg,mahan} that
in the case without AFLRO, this interaction can induce the dressed
holon pairing state (then the electron Cooper pairing state) by
exchanging dressed spin excitations in the higher power of the
hole doping concentration $x$. The angle resolved photoemission
spectroscopy (ARPES) measurements \cite{shen1} have shown that in
the real space the gap function and pairing force have a range of
one lattice spacing, this indicates that the order parameter for
the electron Cooper pair can be expressed as,
\begin{eqnarray}
\Delta&=&\langle C^{\dagger}_{i\uparrow}C^{\dagger}_{i+\hat{\eta}
\downarrow}-C^{\dagger}_{i\downarrow}
C^{\dagger}_{i+\hat{\eta}\uparrow}\rangle \nonumber \\
&=&\langle h_{i\uparrow} h_{i+\hat{\eta}\downarrow}S^{+}_{i}
S^{-}_{i+\hat{\eta}}- h_{i\downarrow}h_{i+\hat{\eta}\uparrow}
S^{-}_{i} S^{+}_{i+\hat{\eta}}\rangle \nonumber \\
&=&-\langle S^{+}_{i} S^{-}_{i+\hat{\eta}}\rangle\Delta_{h},
\end{eqnarray}
with the dressed holon pairing order parameter,
\begin{eqnarray}
\Delta_{h}=\langle h_{i+\hat{\eta}\downarrow}h_{i\uparrow}
-h_{i+\hat{\eta}\uparrow} h_{i\downarrow}\rangle,
\end{eqnarray}
which shows that the SC order parameter is closely related to the
dressed holon pairing amplitude, and is proportional to the number
of doped holes, and not to the number of electrons. Following our
previous discussions \cite{feng1,feng0}, the self-consistent
equations that satisfied by the full dressed holon diagonal and
off-diagonal Green's functions are obtained as,
\begin{mathletters}
\begin{eqnarray}
g(k)&=&g^{(0)}(k)+g^{(0)}(k)[\Sigma^{(h)}_{1}(k)g(k) \nonumber \\
&-&\Sigma^{(h)}_{2}(-k)\Im^{\dagger}(k)], \\
\Im^{\dagger}(k)&=&g^{(0)}(-k)[\Sigma^{(h)}_{1}(-k)
\Im^{\dagger}(-k) \nonumber \\
&+&\Sigma^{(h)}_{2}(-k)g(k)],
\end{eqnarray}
\end{mathletters}
respectively, where the four-vector notation $k=({\bf k},
i\omega_{n})$, and the dressed holon self-energies are obtained
as,
\begin{mathletters}
\begin{eqnarray}
\Sigma^{(h)}_{1}(k)&=&(Zt)^{2}{1\over N^{2}}\sum_{{\bf p,p'}}
\gamma^{2}_{{\bf p+p'+k}}{1\over \beta}\sum_{ip_{m}}g(p+k)
\nonumber \\
&\times&{1\over\beta}\sum_{ip'_{m}}D^{(0)}(p')D^{(0)}(p'+p), \\
\Sigma^{(h)}_{2}(k)&=&(Zt)^{2}{1\over N^{2}}\sum_{{\bf p,p'}}
\gamma^{2}_{{\bf p+p'+k}}{1\over \beta}\sum_{ip_{m}}\Im (-p-k)
\nonumber \\
&\times&{1\over\beta}\sum_{ip'_{m}}D^{(0)}(p')D^{(0)}(p'+p),
\end{eqnarray}
\end{mathletters}
where $p=({\bf p},ip_{m})$, $p'=({\bf p'},ip_{m}')$, and the
dressed holon and spin mean-field (MF) Green's functions are
evaluated as \cite{feng1,feng2,feng0},
\begin{mathletters}
\begin{eqnarray}
g^{(0)}(k)&=& {1\over i\omega_{n}-\xi_{{\bf k}}},\\
D^{(0)}(p)&=&{B_{{\bf p}}\over (ip_{m})^{2}-\omega_{{\bf p}}^{2}},
\end{eqnarray}
\end{mathletters}
with $B_{{\bf p}}=\lambda [2\chi_{z}(\epsilon\gamma_{{\bf p}}-1)+
\chi(\gamma_{{\bf p}}-\epsilon)]$, $\lambda=2ZJ_{{\rm eff}}$,
$\epsilon=1+2t\phi/J_{{\rm eff}}$, $\gamma_{{\bf p}}=(1/Z)
\sum_{\hat{\eta}}e^{i{\bf p}\cdot \hat{\eta}}$, $Z$ is the number
of the nearest neighbor sites, the dressed spin correlation
functions $\chi=\langle S_{i}^{+} S_{i+\hat{\eta}}^{-}\rangle$ and
$\chi_{z}=\langle S_{i}^{z} S_{i+\hat{\eta}}^{z}\rangle$, and the
MF dressed holon and spin excitation spectra are given by,
\begin{mathletters}
\begin{eqnarray}
\xi_{{\bf k}}&=&Zt\chi\gamma_{{\bf k}}-\mu, \\
\omega^{2}_{{\bf p}}&=&\lambda^{2}[(A_{1}-\alpha\epsilon
\chi_{z}\gamma_{{\bf p}}-{1\over 2Z}\alpha\epsilon\chi)
(1-\epsilon\gamma_{{\bf p}}) \nonumber \\
&+&{1\over 2}\epsilon(A_{2}-{1\over 2} \alpha\chi_{z}-
\alpha\chi\gamma_{{\bf p}})(\epsilon-\gamma_{{\bf p}})],
\end{eqnarray}
\end{mathletters}
where $A_{1}=\alpha C^{z}+(1-\alpha)/(4Z)$, $A_{2}=\alpha C+
(1-\alpha)/(2Z)$, the dressed holon particle-hole parameter
$\phi=\langle h^{\dagger}_{i\sigma} h_{i+\hat{\eta}\sigma}
\rangle$, and the dressed spin correlation functions $C=(1/Z^{2})
\sum_{\hat{\eta},\hat{\eta'}}\langle S_{i+\hat{\eta}}^{+}
S_{i+\hat{\eta'}}^{-}\rangle$ and $C_{z}= (1/Z^{2})
\sum_{\hat{\eta},\hat{\eta'}}\langle S_{i+\hat{\eta}}^{z}
S_{i+\hat{\eta'}}^{z}\rangle$. In order to satisfy the sum rule of
the dressed spin correlation function $\langle S^{+}_{i}
S^{-}_{i}\rangle=1/2$ in the case without AFLRO, the important
decoupling parameter $\alpha$ has been introduced in the MF
calculation \cite{feng3,kondo}. In the calculation of the
self-energies (6), the dressed spin part has been limited to the
MF level \cite{feng1}, i.e., the full dressed spin Green's
function in Eq. (6) has been replaced by the MF dressed spin
Green's function (7b), since the normal-state charge transport
obtained at this level can well describe the experimental data
\cite{feng2,feng4}.

Since the pairing force and dressed holon gap function have been
incorporated into the self-energy function $\Sigma^{(h)}_{2}(k)$,
then it is called as the effective dressed holon gap function. On
the other hand, the self-energy function $\Sigma^{(h)}_{1}(k)$
renormalizes the MF dressed holon spectrum, and therefore it
describes the quasiparticle coherence. Moreover,
$\Sigma^{(h)}_{2}(k)$ is an even function of $i\omega_{n}$, while
$\Sigma^{(h)}_{1}(k)$ is not. In this case, it is convenient to
break $\Sigma^{(h)}_{1}(k)$ up into its symmetric and
antisymmetric parts as, $\Sigma^{(h)}_{1}(k)= \Sigma^{(h)}_{1e}(k)
+i\omega_{n} \Sigma^{(h)}_{1o}(k)$, where $\Sigma^{(h)}_{1e}(k)$
and $\Sigma^{(h)}_{1o}(k)$ are both even functions of
$i\omega_{n}$. Now we define the charge carrier quasiparticle
coherent weight $Z^{-1}_{F}(k)=1- \Sigma^{(h)}_{1o}(k)$, then the
dressed holon diagonal and off-diagonal Green's functions in Eq.
(5) can be expressed as,
\begin{mathletters}
\begin{eqnarray}
g(k)&=&{i\omega_{n}/Z_{F}(k)+\xi_{{\bf k}}+\Sigma^{(h)}_{1e}(k)
\over [i\omega_{n}/Z_{F}(k)]^{2}-[\xi_{{\bf k}}+
\Sigma^{(h)}_{1e}(k)]^{2}-[\Sigma^{(h)}_{2}(k)]^{2}} ,\\
\Im^{\dagger}(k)&=&{-\Sigma^{(h)}_{2}(k)\over
[i\omega_{n}/Z_{F}(k)]^{2}-[\xi_{{\bf k}}+\Sigma^{(h)}_{1e}
(k)]^{2}- [\Sigma^{(h)}_{2}(k)]^{2}}.
\end{eqnarray}
\end{mathletters}
As in the conventional superconductor \cite{eliashberg}, the
retarded function ${\rm Re}\Sigma^{(h)}_{1e} (k)$ may be a
constant, independent of (${\bf k},\omega$). It just renormalizes
the chemical potential, and therefore can be neglected.
Furthermore, we only study the static limit of the effective
dressed holon gap function and quasiparticle coherent weight,
i.e., $\Sigma^{(h)}_{2}(k)= \bar{\Delta}_{h}({\bf k})$, and
$Z^{-1}_{F}({\bf k})=1- \Sigma^{(h)}_{1o}({\bf k})$. In this case,
the dressed holon diagonal and off-diagonal Green's functions in
Eq. (9) can be rewritten explicitly as,
\begin{mathletters}
\begin{eqnarray}
g(k)&=&{1\over 2}\left (1+ {\bar{\xi_{{\bf k}}} \over E_{{\bf k}}}
\right ){Z_{F}({\bf k})\over i\omega_{n}-E_{{\bf k}}} \nonumber \\
&+&{1\over 2} \left (1- {\bar{\xi_{{\bf k}}}\over E_{{\bf k}}}
\right ){Z_{F}({\bf k})\over i\omega_{n}+E_{{\bf k}}}, \\
\Im^{\dagger}(k)&=&-{1\over 2}{\bar{\Delta}_{hZ}({\bf k})\over
E_{{\bf k}}}Z_{F}({\bf k})\left ( {1\over i\omega_{n}-E_{{\bf k}}}
\right .\nonumber \\
&-&\left . {1\over i\omega_{n}+ E_{{\bf k}}}\right ),
\end{eqnarray}
\end{mathletters}
with $\bar{\xi_{{\bf k}}}=Z_{F}({\bf k})\xi_{{\bf k}}$, $\bar
{\Delta}_{hZ}({\bf k})=Z_{F}({\bf k})\bar{\Delta}_{h}({\bf k})$,
and the dressed holon quasiparticle spectrum $E_{{\bf k}}=\sqrt
{\bar {\xi}^{2}_{{\bf k}}+\mid\bar{\Delta}_{hZ}({\bf k})\mid^{2}}
$, this $Z_{F}({\bf k})$ reduces the dressed holon quasiparticle
bandwidth. Although $Z_{F}({\bf k})$ is still a function of ${\bf
k}$, the wave vector dependence is unimportant, since everything
happens at the electron Fermi surface (EFS). In this case, we will
approximate $Z_{F}({\bf k})$ by a constant, $Z_{F}=Z_{F}({\bf k}
_{0})$, where the special wave vector ${\bf k}_{0}$ is defined
below. In the CSS fermion-spin theory, the electron diagonal
Green's function $G(i-j,t-t')= \langle\langle C_{i\sigma}(t);
C^{\dagger}_{j\sigma} (t')\rangle \rangle$ is a convolution of the
dressed spin Green's function and dressed holon diagonal Green's
function, which reflects the charge-spin recombination
\cite{anderson3}, and in the present case, it can be calculated in
terms of Eqs. (7b) and (10a) as \cite{feng3},
\begin{eqnarray}
G(k)&=&{1\over N}\sum_{{\bf p}}\int^{\infty}_{-\infty}{d\omega'
\over 2\pi}{d\omega''\over 2\pi}A_{s}({\bf p},\omega') A_{h}({\bf
p-k},\omega'') \nonumber \\
&\times&{n_{F}(\omega'')+n_{B}(\omega')\over i\omega_{n}
+\omega''-\omega'} \nonumber \\
&=&{1\over N}\sum_{{\bf p}}Z_{F}({\bf p-k}){B_{{\bf p}}\over 4
\omega_{{\bf p}}}\left\{\left (1+{\bar{\xi}_{{\bf p-k}}\over
E_{{\bf p-k}}}\right ) \right .\nonumber \\
&\times& \left ({L_{1}({\bf k,p})\over i\omega_{n}+ E_{{\bf p-k}}-
\omega_{{\bf p}}}+{L_{2}({\bf k,p})\over i\omega_{n}
+E_{{\bf p-k}}+\omega_{{\bf p}}}\right ) \nonumber \\
&+&\left . \left (1-{\bar{\xi}_{{\bf p-k}}\over E_{{\bf p-k}}}
\right ) \left ({L_{2}({\bf k,p})\over i\omega_{n}-E_{{\bf p-k}}-
\omega_{{\bf p}}} \right.\right . \nonumber \\
&+&\left. \left . {L_{1}({\bf k,p})\over i\omega_{n}-E_{{\bf p-k}}
+\omega_{{\bf p}}}\right )\right\} ,
\end{eqnarray}
where the MF dressed spin spectral function $A_{s}({\bf k},
\omega) =-2{\rm Im}D^{(0)}({\bf k},\omega)$, the dressed holon
spectral function $A_{h}({\bf k},\omega) =-2{\rm Im}g({\bf k},
\omega)$, $L_{1}({\bf k,p})=n_{F}(E_{{\bf p-k}})+ n_{B}
(\omega_{{\bf p}})$, $L_{2}({\bf k,p})=1-n_{F}(E_{{\bf p-k}})+
n_{B}(\omega_{{\bf p}})$, and $n_{B}(\omega)$ and $n_{F}(\omega)$
are the boson and fermion distribution functions, respectively.
Then the electron quasiparticle dispersion is determined by the
poles of the electron diagonal Green's function (11). At the
half-filling, the $t$-$J$ model is reduced to the AF Heisenberg
model, where there is no charge degree of freedom, and then the
dressed holon excitation spectrum disappears, while the electron
quasiparticle dispersion is reduced as the spin excitation
spectrum \cite{feng9}. This electron diagonal Green's function can
be used to extract the electron momentum distribution (then EFS)
as \cite{feng3},
\begin{eqnarray}
n_{{\bf k}}={1\over 2}-{1\over N}\sum_{{\bf p}}n_{s}({\bf p})
\int^{\infty}_{-\infty}{d\omega\over 2\pi}A_{h}({\bf p-k},
\omega)n_{F}(\omega),
\end{eqnarray}
with $n_{s}({\bf p})=\int^{\infty}_{-\infty}d\omega A_{s}({\bf p},
\omega)n_{s}(\omega)/2\pi$ is the dressed spin momentum
distribution. Then this electron momentum distribution can be
evaluated explicitly in terms of the MF dressed spin Green's
function (7b) and dressed holon diagonal Green's function (10a)
as,
\begin{eqnarray}
n_{{\bf k}}={1\over 2}&-&{1\over 2N}\sum_{{\bf p}}n^{(0)}_{s}({\bf
p})Z_{F}({\bf p-k}) \nonumber \\
&\times& \left (1- {\bar{\xi}_{{\bf p-k}}\over E_{{\bf p-k}}}{\rm
tanh} [{1\over 2}\beta E_{{\bf p-k}}]\right ),
\end{eqnarray}
with $n^{(0)}_{s}({\bf p})=B_{{\bf p}}{\rm coth}(\beta\omega_{{\bf
p}}/2)/(2\omega_{{\bf p}})$. Since the dressed spins center around
$[\pm\pi,\pm\pi]$ in the Brillouin zone at the MF level
\cite{feng3}, then the electron momentum distribution (13) can be
approximately reduced as,
\begin{eqnarray}
n_{{\bf k}}\approx 1/2&-&\rho^{(0)}_{s} Z_{F}({\bf k_{A}-k})
\nonumber \\
&\times& [1-\bar{\xi}_{{\bf k_{A}-k}}{\rm tanh}(\beta E_{{\bf
k_{A}-k}}/2)/ E_{{\bf k_{A}-k}}]/2,
\end{eqnarray}
with ${\bf k_{A}}= [\pi,\pi]$, and $\rho^{(0)}_{s}=(1/N)\sum_{{\bf
p}= (\pm\pi, \pm\pi)}n^{(0)}_{s} ({\bf p})$. It has been shown
from ARPES experiments \cite{shen2} that EFS is small pockets
around $[\pi/2,\pi/2]$ at small doping, and becomes a large EFS at
large doping. Therefore in the present case the Fermi wave vector
from above electron momentum distribution can be estimated
qualitatively \cite{feng3} as ${\bf k_{F}}\approx
[(1-x)\pi/2,(1-x)\pi/2]$, and is evolution with doping. Then the
wave vector ${\bf k}_{0}$ is obtained as ${\bf k}_{0}={\bf k_{A}}
-{\bf k_{F}}$, and we only need to calculate $Z_{F}=Z_{F}({\bf
k}_{0})$ as mentioned above. Since the charge-spin recombination
from the convolution of the dressed spin Green's function and
dressed holon diagonal Green's function leads to form EFS
\cite{anderson3}, then the dressed holon quasiparticle coherence
$Z_{F}$ appearing in the electron momentum distribution also
reflects the electron quasiparticle coherence. We emphasize that
the Fermi wave vector ${\bf k_{F}}$ estimated in the present case
only is qualitative correct, while the quantitative correct EFS
obtained within the $t$-$J$ model is an rather complicated
problem, and one may need to consider the vertex corrections. This
and related theoretical ARPES results are under investigations
now.

Some experiments seem consistent with an s-wave pairing
\cite{chaudhari}, while other measurements gave the evidence in
favor of the d-wave pairing \cite{martindale,tsuei}. This reflects
a fact that the d-wave gap function $\propto {\rm k}^{2}_{x}-{\rm
k}^{2}_{y}$ belongs to the same representation $\Gamma_{1}$ of the
orthorhombic crystal group as does s-wave gap function $\propto
{\rm k}^{2}_{x}+{\rm k}^{2}_{y}$. Within some strong correlated
models, the earlier numerical simulations \cite{scalettar} have
shown that the s-wave channel was competitive with the d-wave,
which indicates that superconductivity with both s-wave and d-wave
symmetries may arise directly from the repulsive interactions. For
understanding of these experimental results, we consider both
s-wave and d-wave cases, i.e., $\bar{\Delta}^{(s)}_{hZ}({\bf k})=
\bar{\Delta}^{(s)}_{hZ}\gamma^{(s)}_{{\bf k}}$, with
$\gamma^{(s)}_{{\bf k}}=\gamma_{{\bf k}}=({\rm cos}k_{x}+{\rm cos}
k_{y})/2$, for the s-wave pairing, and $\bar{\Delta}^{(d)}_{hZ}
({\bf k})=\bar{\Delta}^{(d)}_{hZ}\gamma^{(d)}_{{\bf k}}$, with
$\gamma^{(d)}_{{\bf k}}=({\rm cos} k_{x}-{\rm cos}k_{y})/2$, for
the d-wave pairing, respectively. In this case, the dressed holon
effective gap parameter and quasiparticle coherent weight in Eq.
(6) satisfy the following two equations \cite{feng1,feng0},
\begin{mathletters}
\begin{eqnarray}
1&=&(Zt)^{2}{1\over N^{3}}\sum_{{\bf k,q,p}}\gamma^{2}_{{\bf k+q}}
\gamma^{(a)}_{{\bf k-p+q}}\gamma^{(a)}_{{\bf k}}{Z^{2}_{F}\over
E_{{\bf k}}}{B_{{\bf q}}B_{{\bf p}}\over\omega_{{\bf q}}
\omega_{{\bf p}}} \nonumber \\
&\times& \left({F^{(1)}_{1}({\bf k,q,p})\over (\omega_{{\bf p}}-
\omega_{{\bf q}})^{2}-E^{2}_{{\bf k}}} +{F^{(2)}_{1}({\bf k,q,p})
\over (\omega_{{\bf p}}+\omega_{{\bf q}})^{2}-E^{2}_{{\bf k}}}
\right ) ,\\
Z^{-1}_{F}&=&1+(Zt)^{2}{1\over N^{2}}\sum_{{\bf q,p}}
\gamma^{2}_{{\bf p+k_{0}}}Z_{F}{B_{{\bf q}}B_{{\bf p}}\over
4\omega_{{\bf q}}\omega_{{\bf p}}} \nonumber \\
&\times& \left({F^{(1)}_{2}({\bf q,p}) \over (\omega_{{\bf p}}
-\omega_{{\bf q}}-E_{{\bf p-q+k_{0}}})^{2}}\right . \nonumber \\
&+&{F^{(2)}_{2}({\bf q,p})\over (\omega_{{\bf p}}- \omega_{{\bf
q}} +E_{{\bf p-q+k_{0}}})^{2}} \nonumber \\
&+&{F^{(3)}_{2}({\bf q,p}) \over (\omega_{{\bf p}}+ \omega_{{\bf
q}}-E_{{\bf p-q+k_{0}}} )^{2}} \nonumber \\
&+&\left . {F^{(4)}_{2}({\bf q,p})\over (\omega_{{\bf
p}}+\omega_{{\bf q}}+E_{{\bf p-q+k_{0}}})^{2}} \right ) ,
\end{eqnarray}
\end{mathletters}
respectively, where $a={\rm s,d}$, $F^{(1)}_{1}({\bf k,q,p})=
(\omega_{{\bf p}}- \omega_{{\bf q}})[n_{B}(\omega_{{\bf q}})-
n_{B}(\omega_{{\bf p}})] [1-2 n_{F}(E_{{\bf k}})]+E_{{\bf k}}
[n_{B}(\omega_{{\bf p}}) n_{B}( -\omega_{{\bf q}})+
n_{B}(\omega_{{\bf q}}) n_{B}(-\omega_{{\bf p}}) ]$,
$F^{(2)}_{1}({\bf k,q,p}) =-(\omega_{{\bf p }}+\omega_{{\bf q}})
[n_{B}(\omega_{{\bf q}}) -n_{B}(-\omega_{{\bf p}})][1-2 n_{F}
(E_{{\bf k}})]+E_{{\bf k}} [n_{B}(\omega_{{\bf p}}) n_{B}
(\omega_{{\bf q}}) +n_{B}(-\omega_{{\bf p}})n_{B}(-\omega_{{\bf q}
})]$, $F^{(1)}_{2}({\bf q,p})=n_{F}(E_{{\bf p- q+k_{0}}})[n_{B}
(\omega_{{\bf q}})-n_{B}(\omega_{{\bf p}})]- n_{B}(\omega_{{\bf p}
})n_{B}(-\omega_{{\bf q}})$, $F^{(2)}_{2} ({\bf q,p})= n_{F}
(E_{{\bf p-q+k_{0}}}) [n_{B}(\omega_{{\bf p}})-n_{B} (\omega_{{\bf
q}})]-n_{B}(\omega_{{\bf q}})n_{B} (-\omega_{{\bf p}})$,
$F^{(3)}_{2}({\bf q,p})= n_{F}(E_{{\bf p-q+k_{0}}})
[n_{B}(\omega_{{\bf q}})-n_{B}(-\omega_{{\bf p}})]+n_{B}
(\omega_{{\bf p}})n_{B}(\omega_{{\bf q}})$, and $F^{(4)}_{2}({\bf
q,p})=n_{F}(E_{{\bf p-q+k_{0}}})[n_{B}(-\omega_{{\bf q}})-n_{B}
(\omega_{{\bf p}})]+n_{B}(-\omega_{{\bf p}})n_{B}(-\omega_{{\bf
q}})$. These two equations are in control of the SC order
directly, and must be solved simultaneously with other
self-consistent equations as shown in Ref. \cite{feng1}, then all
order parameters, decoupling parameter $\alpha$, and chemical
potential $\mu$ are determined by the self-consistent calculation
\cite{feng1}. In this case, we obtain the dressed holon pair gap
function in terms of the off-diagonal Green's function (10b) as
\cite{eliashberg},
\begin{eqnarray}
\Delta^{(a)}_{h}({\bf k})&=&-{1\over \beta}\sum_{i\omega_{n}}
\Im^{\dagger}({\bf k},i\omega_{n}) \nonumber \\
&=&{1\over 2}Z_{F} {\bar{\Delta}^{(a)}_{hZ}({\bf k})\over E_{{\bf
k}}}{\rm tanh} [{1\over 2}\beta E_{{\bf k}}],
\end{eqnarray}
then the dressed holon pair order parameter in Eq. (4) can be
evaluated explicitly as,
\begin{eqnarray}
\Delta^{(a)}_{h}={2\over N}\sum_{{\bf k}}[\gamma^{(a)}_{{\bf k}}
]^{2}{Z_{F}\bar{\Delta}^{(a)}_{hZ}\over E_{{\bf k}}}{\rm tanh}
[{1\over 2}\beta E_{{\bf k}}].
\end{eqnarray}
We \cite{feng1,feng0} have shown that this dressed holon pairing
state originating from the kinetic energy term by exchanging
dressed spin excitations can lead to form the electron Cooper
pairing state, where the SC gap function is obtained from the
electron off-diagonal Green's function $I^{\dagger}(i-j,t-t')=
\langle \langle C^{\dagger}_{i\uparrow}(t);
C^{\dagger}_{j\downarrow} (t')\rangle \rangle$, which is a
convolution of the dressed spin Green's function and dressed holon
off-diagonal Green's function \cite{anderson3}, and in the present
case can be obtained in terms of the dressed spin MF Green's
function (7b) and dressed holon off-diagonal Green's function
(10b) as,
\begin{eqnarray}
I^{\dagger}(k)&=&{1\over N}\sum_{{\bf p}} {Z_{F}
\bar{\Delta}^{(a)}_{hZ}({\bf p-k})\over E_{{\bf p-k}}}{B_{{\bf p}}
\over 2\omega_{{\bf p}}}\nonumber\\
&\times&\left ({(\omega_{{\bf p}}+E_{{\bf p-k}})
[n_{B}(\omega_{{\bf p}})+n_{F}(-E_{{\bf p-k}})] \over
(i\omega_{n})^{2}-(\omega_{{\bf p}}+E_{{\bf p-k}})^{2}}
\right . \nonumber \\
&-& \left . {(\omega_{{\bf p}}-E_{{\bf p-k}})[n_{B}(\omega_{{\bf
p}})+n_{F}(E_{{\bf p-k}})]\over (i\omega_{n})^{2}-(\omega_{{\bf
p}}-E_{{\bf p-k}})^{2}} \right ),
\end{eqnarray}
then the SC gap function is obtained from this electron
off-diagonal Green's function as,
\begin{eqnarray}
\Delta^{(a)}({\bf k})&=&-{1\over \beta}\sum_{i\omega_{n}}
I^{\dagger}({\bf k},i\omega_{n})\nonumber \\
&=&-{1\over N}\sum_{{\bf p}} {Z_{F}\bar{\Delta}^{(a)}_{Zh}({\bf
p-k})\over 2E_{{\bf p-k}}}{\rm tanh}[{1\over 2}\beta E_{{\bf
p-k}}]\nonumber \\
&\times&{B_{{\bf p}}\over 2 \omega_{{\bf p}}}{\rm coth}[{1\over
2}\beta\omega_{{\bf p}}],
\end{eqnarray}
which shows that the symmetry of the electron Cooper pair is
determined by the symmetry of the dressed holon pair \cite{feng1},
and therefore the SC gap function can be written as $\Delta^{(a)}
({\bf k})=\Delta^{(a)} \gamma^{(a)}_{{\bf k}}$, then the SC gap
parameter in Eq. (3) is evaluated in terms of Eqs. (19) and (17)
as $\Delta^{(a)}=-\chi \Delta^{(a)}_{h}$. Since the dressed holon
(then electron) pairing interaction also is doping dependent, then
the experimental observed SC gap parameter should be the effective
SC gap parameter $\bar{\Delta}^{(a)}\sim-\chi\bar
{\Delta}^{(a)}_{h}$. In Fig. 1, we plot the effective dressed
holon pairing (a) and effective SC (b) gap parameters in the
s-wave symmetry (solid line) and d-wave symmetry (dashed line) as
a function of the hole doping concentration $x$ at $T=0.002J$ and
$t/J=2.5$. For comparison, the experimental result \cite{wen} of
the upper critical field as a function of the hole doping
concentration is also shown in Fig. 1(b). In a given doping
concentration, the upper critical field is defined as the critical
field that destroys the SC-state at the zero temperature in the
given doping concentration, therefore the upper critical field
also measures the strength of the binding of electrons into Cooper
pairs like the effective SC gap parameter \cite{wen}. In other
words, both effective SC gap parameter and upper critical field
have a similar doping dependence \cite{wen}. In this sense, our
result is in qualitative agreement with the experimental data
\cite{wen}. In particular, the value of $\bar{\Delta}^{(d)}$
increases with increasing doping in the underdoped regime, and
reaches a maximum in the optimal doping $x_{{\rm opt}}\approx
0.18$, then decreases in the overdoped regime.

\begin{figure}[prb]
\epsfxsize=3.5in\centerline{\epsffile{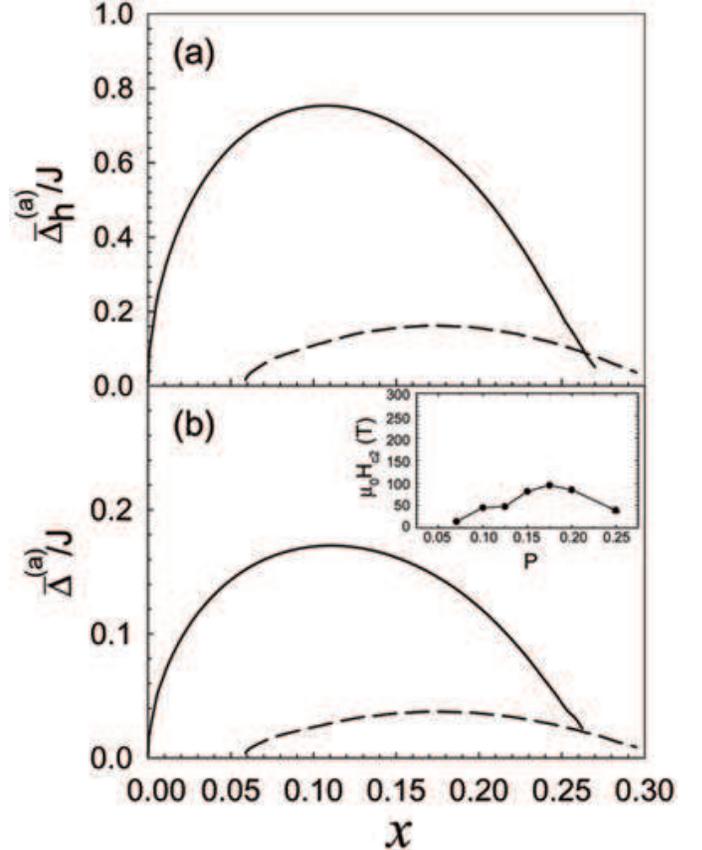}}\caption{The
effective dressed holon pairing (a) and effective superconducting
(b) gap parameters in the s-wave symmetry (solid line) and d-wave
symmetry (dashed line) as a function of the hole doping
concentration in $T=0.002J$ and $t/J=2.5$. Inset: the experimental
result of the upper critical field as a function of the hole
doping concentration taken from Ref. [32].}
\end{figure}

\begin{figure}[prb]
\epsfxsize=3.5in\centerline{\epsffile{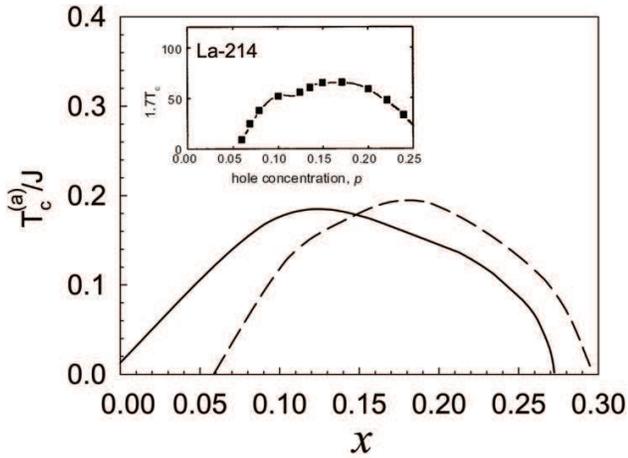}} \caption{The
superconducting transition temperature as a function of the hole
doping concentration in the s-wave symmetry (solid line) and
d-wave symmetry (dashed line) for $t/J=2.5$. Inset: the
experimental result taken from Ref. [5].}
\end{figure}

\begin{figure}[prb]
\epsfxsize=3.5in\centerline{\epsffile{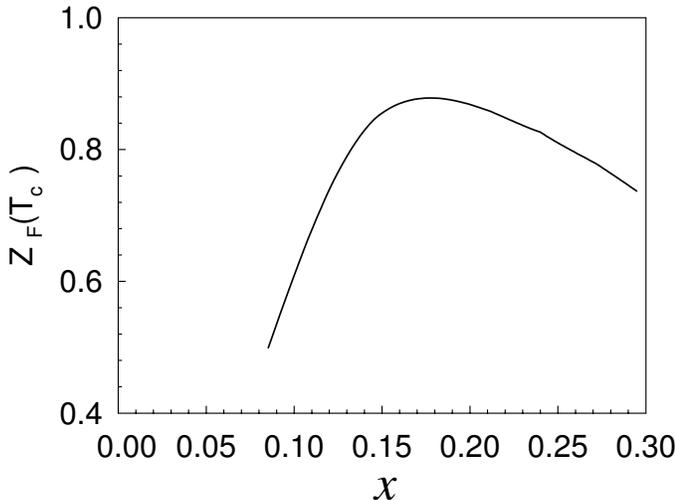}} \caption{The
quasiparticle coherent weight $Z_{F}(T_{c})$ as a function of the
hole doping concentration for $t/J=2.5$.}
\end{figure}

The present result in Eq. (19) also shows that the SC transition
temperature $T^{(a)}_{c}$ occurring in the case of the SC gap
parameter $\Delta^{(a)}=0$ is identical to the dressed holon pair
transition temperature occurring in the case of the effective
dressed holon pairing gap parameter $\bar{\Delta}^{(a)}_{h}=0$. In
correspondence with the SC gap parameter, the SC transition
temperature $T^{(a)}_{c}$ as a function of the hole doping
concentration $x$ in the s-wave (solid line) and d-wave (dashed
line) symmetries for $t/J=2.5$ is plotted in Fig. 2 in comparison
with the experimental result \cite{tallon} (inset). For the s-wave
symmetry, the maximal SC transition temperature T$^{(s)}_{c}$
occurs around a particular doping concentration $x\approx 0.11$,
and then decreases in both lower doped and higher doped regimes.
However, for the d-wave symmetry, the maximal SC transition
temperature T$^{(d)}_{c}$ occurs around the optimal doping
concentration $x_{{\rm opt}}\approx 0.18$, and then decreases in
both underdoped and overdoped regimes. Although the SC pairing
symmetry is doping dependent, the SC state has the d-wave symmetry
in a wide range of doping, in qualitative agreement with the
experiments \cite{yeh,biswas,tsuei1}. Furthermore, T$^{(d)}_{c}$
in the underdoped regime (T$^{(s)}_{c}$ in the lower doped regime)
is proportional to the hole doping concentration $x$, and
therefore T$^{(d)}_{c}$ in the underdoped regime (T$^{(s)}_{c}$ in
the lower doped regime) is set by the hole doping concentration
\cite{uemura}, this reflects that the density of the dressed
holons directly determines the superfluid density in the
underdoped regime for the d-wave case (the lower doped regime for
the s-wave case). Using an reasonably estimative value of $J\sim
800$K to 1200K in doped cuprates, the SC transition temperature in
the optimal doping is T$^{(d)}_{c} \approx 0.2J \approx 160{\rm
K}\sim 240{\rm K}$, also in qualitative agreement with the
experimental data \cite{tallon,uemura,tsuei1}.

In the present framework of the kinetic energy driven
superconductivity, the antisymmetric part of the self-energy
function $\Sigma^{(h)}_{1o}({\bf k})$ (then $Z_{F}$) describes the
quasiparticle coherence, and therefore $Z_{F}$ is closely related
to the quasiparticle density, while the self-energy function
$\Sigma^{(h)}_{2}({\bf k})$ describes the effective dressed holon
pairing gap function. Since the SC-order is established through an
emerging quasiparticle \cite{ding}, therefore the SC-order is
controlled by both gap function and quasiparticle coherence, and
is reflected explicitly in the self-consistent equations (15a) and
(15b). To show this point clearly, we plot the quasiparticle
coherent weight $Z_{F}(T_{c})$ as a function of the hole doping
concentration $x$ for $t/J=2.5$ in Fig. 3. As seen from Fig. 3,
the doping dependent behavior of the quasiparticle coherent weight
resembles that of the superfluid density in doped cuprates, i.e.,
$Z_{F}$ grows linearly with the hole doping concentration in the
underdoped and optimally doped regimes, and then decreases with
increasing doping in the overdoped regime, which leads to that the
SC transition temperature reaches a maximum in the optimal doping,
and then decreases in both underdoped and overdoped regimes. In
comparison with Ref. \cite{feng1}, we therefore find that the
quasiparticle coherence plays an important role in the kinetic
energy driven superconductivity of doped cuprates. Since cuprate
superconductors are highly anisotropic materials, therefore the
effective SC gap function $\bar{\Delta}^{(s)}({\bf k})=
\bar{\Delta}^{(s)}({\rm cos}k_{x}+{\rm cos} k_{y})/2$ for the
s-wave symmetry or $\bar{\Delta}^{(d)}({\bf k})=
\bar{\Delta}^{(d)}({\rm cos}k_{x}-{\rm cos}k_{y})/2$ for the
d-wave case is dependent on the momentum. According to a
comparison of the density of states as measured by scanning
tunnelling microscopy \cite{dewilde} and ARPES spectral function
\cite{ding} at $[\pi,0]$ point on identical samples, it has been
shown that the most contributions of the electronic states come
from $[\pi,0]$ point. In this case, although the value of the
effective SC gap parameter $\bar{\Delta}^{(s)}$ (then the ratio
$\bar{\Delta}^{(s)}$/T$^{(s)}_{c}$) for the s-wave symmetry is
larger than these $\bar{\Delta}^{(d)}$ (then the ratio
$\bar{\Delta}^{(d)}$/T$^{(d)}_{c}$) in the d-wave case, the system
has the SC transition temperature T$^{(d)}_{c}$ with the d-wave
symmetry in a wide range of doping.

\section{Doping and energy dependent magnetic excitations}

In the CSS fermion-spin theory, the AF fluctuation is dominated by
the scattering of the dressed spins \cite{feng2,feng5}. Since in
the normal-state the dressed spins move in the dressed holon
background, therefore the dressed spin self-energy (then full
dressed spin Green's function) in the normal-state has been
obtained in terms of the collective mode in the dressed holon
particle-hole channel \cite{feng2,feng5}. With the help of this
full dressed spin Green's function in the normal-state, the IC
magnetic scattering and integrated spin response of doped cuprates
in the normal-state have been discussed \cite{feng2,feng5}, and
the results of the doping dependence of the incommensurability and
integrated dynamical spin susceptibility are consistent with
experimental results in the normal-state \cite{kastner,yamada}.
However, in the present SC-state discussed in Sec. II, the AF
fluctuation has been incorporated into the electron off-diagonal
Green's function (18) (hence the electron Cooper pair) in terms of
the dressed spin Green's function, therefore there is a
coexistence of the electron Cooper pair and AFSRC, and then AFSRC
can persist into superconductivity \cite{feng1}. Moreover, in the
SC-state, the dressed spins move in the dressed holon pair
background. In this case, we calculate the dressed spin
self-energy (then the full dressed spin Green's function) in the
SC-state in terms of the collective mode in the dressed holon
particle-particle channel, and then give a theoretical explanation
of the IC magnetic scattering peaks at both low and high energies
and commensurate resonance peak at intermediate energy in the
SC-state
\cite{dai,yamada,bourges0,bourges,arai,hayden,stock,tranquada}.

Following our previous discussions for the normal-state case
\cite{feng2,feng5}, the full dressed spin Green's functions is
expressed as,
\begin{eqnarray}
D({\bf k},\omega)={1\over D^{(0)-1}({\bf k},\omega)-\Sigma^{(s)}
({\bf k},\omega)},
\end{eqnarray}
with the second order spin self-energy $\Sigma^{(s)}({\bf k},
\omega)$. Within the framework of the equation of motion method
\cite{feng2,feng5}, this self-energy in the SC-state with the
d-wave symmetry is obtained from the dressed holon bubble in the
dressed holon particle-particle channel as,
\begin{eqnarray}
\Sigma^{(s)}(k)&=&(Zt)^{2}{1\over N^{2}}\sum_{{\bf p,p'}}
(\gamma^{2}_{{\bf p'+p+k}}+\gamma^{2}_{{\bf p-k}})\nonumber \\
&\times&{1\over \beta}\sum_{ip_{m}'}D^{(0)}(p'+k)
{1\over\beta}\sum_{ip_{m}} \Im^{\dagger}(p)\Im(p+p'),
\end{eqnarray}
and can be evaluated explicitly in terms of the dressed holon
off-diagonal Green's function (10b) and dressed spin MF Green's
function (7b) as,
\begin{eqnarray}
\Sigma^{(s)}({\bf k},\omega)&=&(Zt)^{2}{1\over N^{2}}\sum_{{\bf
p,q}}(\gamma^{2}_{{\bf q+p+k}}+\gamma^{2}_{{\bf p-k}})\nonumber\\
&\times&{B_{{\bf q+k}}\over\omega_{{\bf q+k}}}{Z^{2}_{F}\over
4}{\bar{\Delta}^{(d)}_{hZ}({\bf p}) \bar{\Delta}^{(d)}_{hZ}({\bf
p+q})\over E_{{\bf p}}E_{{\bf p+q}}} \nonumber \\
&\times& \left ( {F^{(1)}_{s}({\bf k,p,q})\over \omega^{2}-
(E_{{\bf p}} -E_{{\bf p+q}}+\omega_{{\bf q+k}})^{2}} \right .
\nonumber \\
&+&{F^{(2)}_{s}({\bf k,p,q})\over \omega^{2}-(E_{{\bf p+q}}
-E_{{\bf p}}+\omega_{{\bf q+k}})^{2}} \nonumber \\
&+& {F^{(3)}_{s}({\bf k,p,q})\over \omega^{2}-(E_{{\bf p}} +
E_{{\bf p+q}} +\omega_{{\bf q+k}})^{2}} \nonumber \\
&+& \left . {F^{(4)}_{s}({\bf k,p,q})\over \omega^{2}- (E_{{\bf
p+q}}+E_{{\bf p}}-\omega_{{\bf q+k}})^{2}} \right ),
\end{eqnarray}
where $F^{(1)}_{s}({\bf k,p,q})=(E_{{\bf p}}-E_{{\bf p+q}}+
\omega_{{\bf q+k}})\{n_{B}(\omega_{{\bf q+k}})[n_{F}(E_{{\bf
p}})-n_{F}(E_{{\bf p+q}})]-n_{F}(E_{{\bf p+q}})n_{F}(-E_{{\bf
p}})\}$, $F^{(2)}_{s}({\bf k,p,q})=(E_{{\bf p+q}}-E_{{\bf p}}+
\omega_{{\bf q+k}})\{n_{B} (\omega_{{\bf q+k}})[n_{F}(E_{{\bf
p+q}})-n_{F} (E_{{\bf p}})]-n_{F}(E_{{\bf p}})n_{F}(-E_{{\bf
p+q}})\}$, $F^{(3)}_{s}({\bf k,p,q}) =(E_{{\bf p}}+E_{{\bf p+q}}
+\omega_{{\bf q+k}})\{n_{B} (\omega_{{\bf q+k}})[n_{F}(-E_{{\bf
p}})- n_{F}(E_{{\bf p+q}})] +n_{F}(-E_{{\bf p+q}})n_{F}(-E_{{\bf
p}})\}$, $F^{(4)}_{s}({\bf k,p,q})=(E_{{\bf p}}+E_{{\bf p+q}}-
\omega_{{\bf q+k}})\{n_{B} (\omega_{{\bf q+k}})[n_{F}(-E_{{\bf
p}})-n_{F}(E_{{\bf p+q}})] -n_{F}(E_{{\bf p+q}})n_{F}(E_{{\bf
p}})\}$. With the help of this full dressed spin Green's function,
we can obtain the dynamical spin structure factor in the SC-state
with the d-wave symmetry as,
\begin{eqnarray}
&S&({\bf k},\omega)=-2[1+n_{B}(\omega)]{\rm Im}D({\bf k},
\omega)= 2[1+n_{B}(\omega)]\nonumber \\
&\times& {B^{2}_{{\bf k}}{\rm Im} \Sigma^{(s)}({\bf k},\omega)
\over [\omega^{2}-\omega^{2}_{{\bf k}}-B_{{\bf k}}{\rm Re}
\Sigma^{(s)}({\bf k},\omega)]^{2}+[B_{{\bf k}}{\rm Im}
\Sigma^{(s)}({\bf k},\omega)]^{2}},
\end{eqnarray}
where ${\rm Im}\Sigma^{(s)}({\bf k},\omega)$ and ${\rm Re}
\Sigma^{(s)}({\bf k},\omega)$ are the imaginary and real parts of
the second order spin self-energy in Eq. (18), respectively.

\begin{figure}[prb]
\epsfxsize=3.5in\centerline{\epsffile{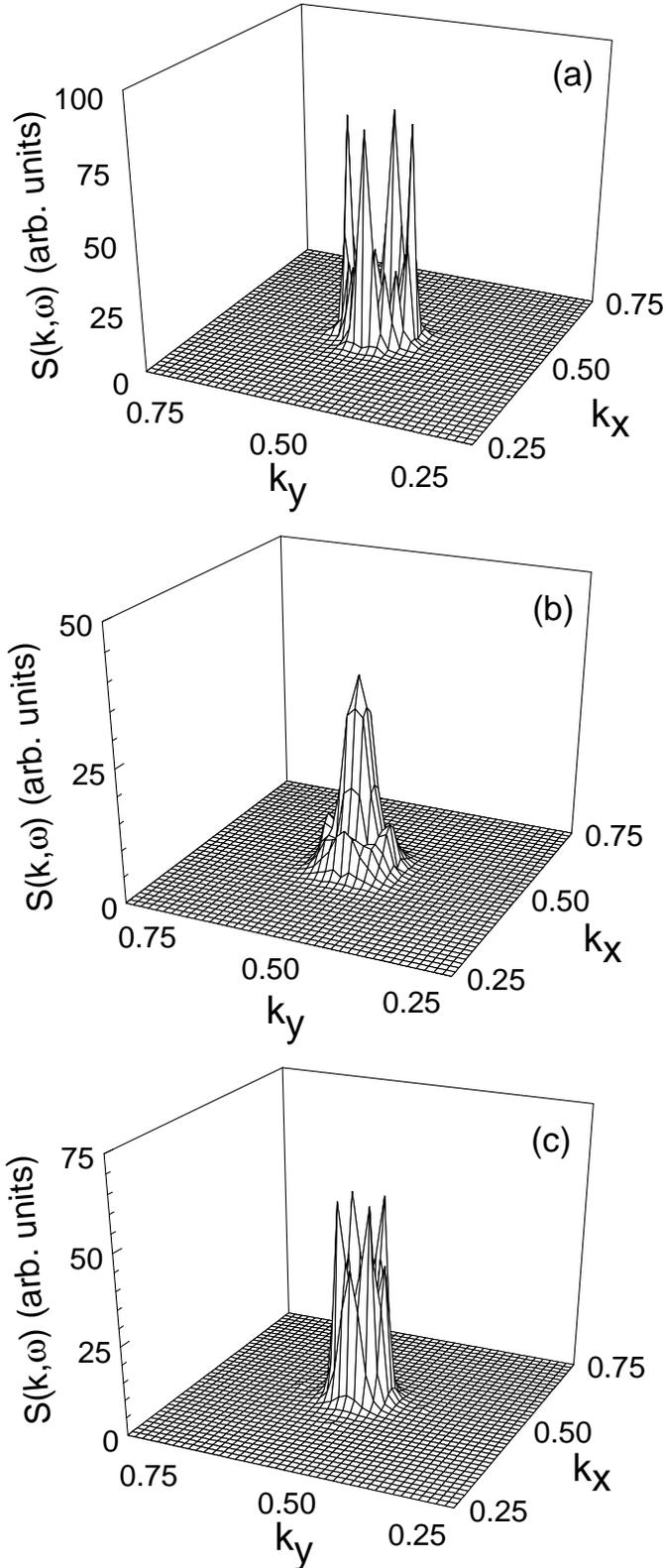}} \caption{The
dynamical spin structure factor $S({\bf k},\omega)$ in the
($k_{x},k_{y}$) plane at $x_{{\rm opt}}=0.18$ with $T=0.002J$ for
$t/J=2.5$ at (a) $\omega =0.13J$, (b) $\omega =0.35J$, and (c)
$\omega =0.65J$.}
\end{figure}

\begin{figure}[prb]
\epsfxsize=3.5in\centerline{\epsffile{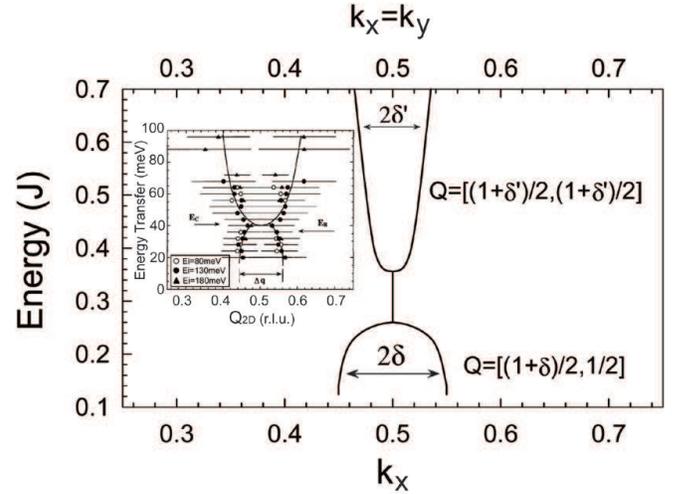}} \caption{The
energy dependence of the position of the magnetic scattering peaks
at $x_{{\rm opt}}=0.18$ and $T=0.002J$ for $t/J=2.5$. Inset: the
experimental result on YBa$_{2}$Cu$_{3}$O$_{6.85}$ in the
superconducting-state taken from Ref. [12]}
\end{figure}

\begin{figure}[prb]
\epsfxsize=3.5in\centerline{\epsffile{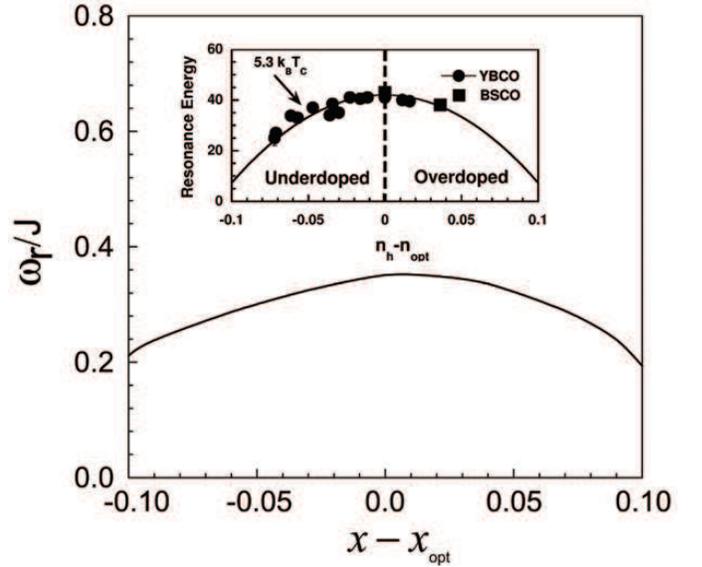}} \caption{The
resonance energy $\omega_{r}$ as a function of $x-x_{{\rm opt}}$
in $T=0.002J$ for $t/J=2.5$. Inset: the experimental result taken
from Ref. [10].}
\end{figure}

\begin{figure}[prb]
\epsfxsize=3.5in\centerline{\epsffile{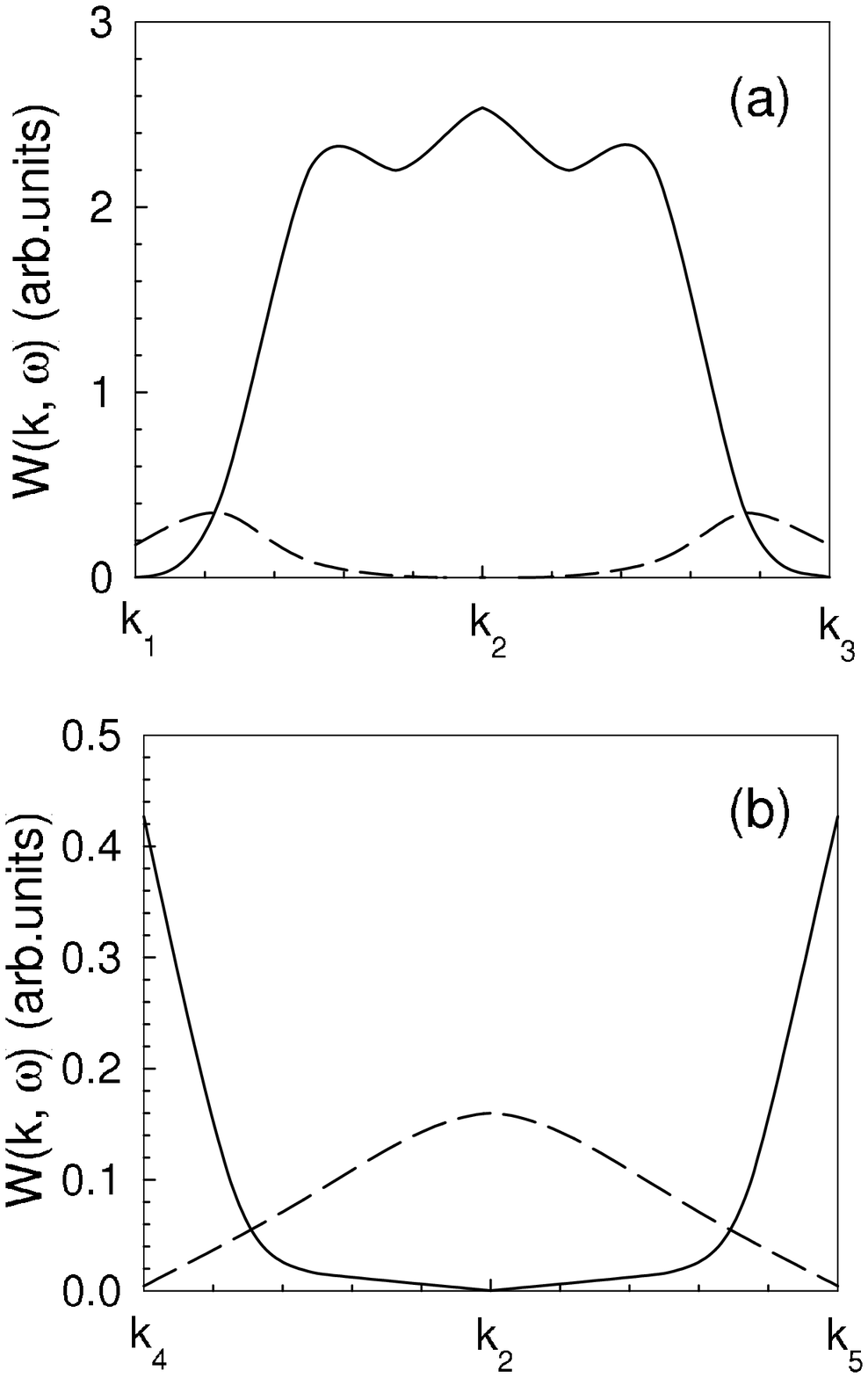}} \caption{Function
$W({\bf k},\omega)$ in $x_{{\rm opt}}=0.18$ for $t/J=2.5$ with
$T=0.002J$ from (a) ${\bf k}_{1}=[(1-\delta)/2,1/2] $ via ${\bf
k}_{2}=[1/2,1/2]$ to ${\bf k}_{3}=[(1+\delta)/2,1/2]$ at
$\omega=0.13J$ (solid line) and $\omega=0.35J$ (dashed line), and
(b) ${\bf k}_{4}=[(1-\delta')/2,(1-\delta')/2]$ via ${\bf k}_{2}=
[1/2,1/2]$ to ${\bf k}_{5}=[(1+\delta')/2,(1+\delta')/2]$ at
$\omega=0.35J$ (solid line) and $\omega=0.65J$ (dashed line).}
\end{figure}

We are now ready to discuss the doping and energy dependent
magnetic excitations in the SC-state. In Fig. 4, we plot the
dynamical spin structure factor $S({\bf k},\omega)$ in the
($k_{x},k_{y}$) plane at the optimal doping $x_{{\rm opt}}=0.18$
with temperature $T=0.002J$ for parameter $t/J=2.5$ at energy (a)
$\omega =0.13J$, (b) $\omega =0.35J$, and (c) $\omega =0.65J$,
where the distinct feature is the presence of the
IC-commensurate-IC transition in the spin fluctuation geometry. At
low energy, the IC peaks are located at $[(1\pm\delta)/2,1/2]$ and
$[1/2,(1\pm\delta )/2]$ (hereafter we use the units of $[2\pi
,2\pi ]$). However, these IC peaks are energy dependent, i.e.,
although these magnetic scattering peaks retain the IC pattern at
$[(1\pm \delta)/2,1/2]$ and $[1/2,(1\pm \delta )/2]$ at low
energy, the positions of the IC peaks move towards $[1/2,1/2]$
with increasing energy, and then the commensurate $[1/2,1/2]$
resonance peak appears at intermediate energy $\omega_{r}=0.35J$.
This anticipated resonance energy $\omega_{r}=0.35J\approx 35$ mev
(Ref. \cite{shamoto}) is not too far from the resonance energy
$\approx 41$ mev observed in optimally doped
YBa$_2$Cu$_3$O$_{6+y}$ \cite{dai,bourges0,bourges,arai}.
Furthermore, the IC peaks are separated again above the resonance
energy, and all IC peaks lie on a circle of radius of $\delta'$.
The values of $\delta'$ at high energy are different from the
corresponding values of $\delta$ at low energy. Although some IC
satellite parallel peaks appear, the main weight of the IC peaks
is in the diagonal direction. Moreover, the separation at high
energy gradually increases with increasing energy although the
peaks have a weaker intensity than those below the resonance
energy. To show this point clearly, we plot the evolution of the
magnetic scattering peaks with energy at $x_{{\rm opt}}=0.18$ in
Fig. 5. For comparison, the experimental result \cite{arai} of
YBa$_2$Cu$_3$O$_{6+y}$ with $y=0.7$ $(x\approx 0.12)$ in the
SC-state is shown in the same figure. The similar experimental
results \cite{bourges,hayden} have also been obtained for
YBa$_2$Cu$_3$O$_{6+y}$ with different doping concentrations. Our
results show that there is a narrow energy range for the resonance
peak, and therefore the dispersion at high energy is distinctly
separated from the low energy IC fluctuations. The similar narrow
energy range for the resonance peak has been observed from
experiments \cite{bourges}. The present results also show that in
contrast to the case at low energy, the magnetic excitations at
high energy disperse almost linearly with energy. Furthermore, the
resonance energy $\omega_{r}$ as a function of doping $x-x_{{\rm
opt}}$ in $T=0.002J$ for $t/J=2.5$ is plotted in Fig. 6 in
comparison with the experimental result \cite{bourges0} (inset).
It is shown that in analogy to the doping dependence of the SC
transition temperature, the magnetic resonance energy $\omega_{r}$
increases with increasing doping in the underdoped regime, and
reaches a maximum in the optimal doping, then decreases in the
overdoped regime. These mediating dressed spin excitations in the
SC-state are coupled to the conducting dressed holons (then
electrons) under the kinetic energy driven SC mechanism
\cite{feng1}, and have energy greater than the dressed holon
pairing energy (then Cooper pairing energy). We have also made a
series of scans for $S({\bf k}, \omega)$ at different
temperatures, and found that those unusual magnetic excitations
are present near the SC transition temperature. Although the
simple $t$-$J$ model can not be regarded as a comprehensive model
for the quantitative comparison with cuprate superconductors, our
these results are in qualitative agreement with the major
experimental observations of doped cuprates in the SC-state
\cite{dai,bourges0,bourges,arai,hayden,stock,tranquada}.

The physical interpretation to the above obtained results can be
found from the property of the renormalized dressed spin
excitation spectrum $\Omega^{2}_{{\bf k}}=\omega^{2}_{{\bf k}}+
{\rm Re}\Sigma^{(s)}({\bf k},\Omega_{{\bf k}})$ in Eq. (23). Since
both MF dressed spin excitation spectrum $\omega_{{\bf k}}$ and
dressed spin self-energy function $\Sigma^{(s)}({\bf k},\omega)$
in Eq. (22) are strong doping and energy dependent, this leads to
that the renormalized dressed spin excitation spectrum also is
strong doping and energy dependent. The dynamical spin structure
factor in Eq. (23) has a well-defined resonance character, where
$S({\bf k},\omega)$ exhibits peaks when the incoming neutron
energy $\omega$ is equal to the renormalized spin excitation,
i.e.,
\begin{eqnarray}
W({\bf k}_{c},\omega)&\equiv& [\omega^{2}-\omega_{{\bf k}_{c}}^{2}
-B_{{\bf k}_{c}}{\rm Re}\Sigma^{(s)}({\bf k}_{c}, \omega)]^{2}
\nonumber \\
&=& [\omega^{2}-\Omega_{{\bf k}_{c}}^{2}]^{2}\sim 0,
\end{eqnarray}
for certain critical wave vectors ${\bf k}_{c}={\bf k}^{(L)}_{c}$
at low energy, ${\bf k}_{c}={\bf k}^{(I)}_{c}$ at intermediate
energy, and ${\bf k}_{c}={\bf k}^{(H)}_{c}$ at high energy, then
the weight of these peaks is dominated by the inverse of the
imaginary part of the dressed spin self-energy $1/{\rm Im}
\Sigma^{(s)}({\bf k}^{(L)}_{c},\omega)$ at low energy, $1/{\rm Im}
\Sigma^{(s)}({\bf k}^{(I)}_{c},\omega)$ at intermediate energy,
and $1/{\rm Im}\Sigma^{(s)}({\bf k}^{(H)}_{c},\omega)$ at high
energy, respectively. In the normal-state \cite{feng2,feng5}, the
dressed holon energy spectrum has one branch $\xi_{{\bf k}}$,
while in the present SC-state, the dressed holon quasiparticle
spectrum has two branches $\pm E_{{\bf k}}$, this leads to that
the dressed spin self-energy function $\Sigma^{(s)}({\bf k},
\omega)$ in Eq. (22) is rather complicated, where there are four
terms in the right side of Eq. (22). In comparison with the
normal-state case \cite{feng2,feng5}, the contribution for the
first and second terms in the right side of the dressed spin
self-energy (22) comes from the lower band $-E_{{\bf k}}$ of the
dressed holon quasiparticle spectrum like the normal-state case,
while the contribution for the third and fourth terms in the right
side of the dressed spin self-energy (22) comes from the upper
band $E_{{\bf k}}$ of the dressed holon quasiparticle spectrum.
During the above calculation, we find that the mode which opens
downward and gives the IC magnetic scattering at low energy is
mainly determined by the first and second terms in the right side
of the dressed spin self-energy (22), while the mode which opens
upward and gives the IC magnetic scattering at high energy is
essentially dominated by the third and fourth terms in the right
side of the dressed spin self-energy (22), then two modes meet at
the commensurate $[1/2,1/2]$ resonance at intermediate energy.
This means that within the framework of the kinetic energy driven
superconductivity, as a result of self-consistent motion of the
dressed holon pairs and spins, the IC magnetic scattering at both
low and high energies and commensurate resonance at intermediate
energy are developed. This reflects that the low and high energy
spin excitations drift away from the AF wave vector, or the zero
point of $W({\bf k}_{c},\omega)$ is shifted from $[1/2,1/2]$ to
${\bf k}_{c}={\bf k}^{(L)}_{c}$ at low energy and ${\bf k}_{c}=
{\bf k}^{(H)}_{c}$ at high energy. With increasing energy from low
energy or decreasing energy from high energy, the spin excitations
move towards to $[1/2,1/2]$, i.e., the zero point of $W({\bf
k}_{c},\omega)$ in ${\bf k}_{c}={\bf k}^{(L)}_{c}$ at low energy
or ${\bf k}_{c}={\bf k}^{(H)}_{c}$ at high energy turns back to
$[1/2,1/2]$, then the commensurate $[1/2,1/2]$ resonance appears
at intermediate energy. To show this point clearly, the function
$W({\bf k},\omega)$ in $x_{{\rm opt}} =0.18$ for $t/J=2.5$ with
$T=0.002J$ from (a) ${\bf k}_{1} =[(1- \delta)/2, 1/2]$ via ${\bf
k}_{2}=[1/2,1/2]$ to ${\bf k}_{3}=[(1+\delta)/2,1/2]$ at
$\omega=0.13J$ (solid line) and $\omega= 0.35J$ (dashed line), and
(b) ${\bf k}_{4}=[(1-\delta')/2,(1-\delta')/2]$ via ${\bf k}_{2}=
[1/2,1/2]$ to ${\bf k}_{5}=[(1+\delta')/2,(1+\delta')/2]$ at
$\omega=0.35J$ (solid line) and $\omega=0.65J$ (dashed line) is
plotted in Fig. 7, where there is a strong angular dependence with
actual minima in $[(1-\delta)/2,1/2]$ and $[1/2,(1-\delta)/2]$,
$[1/2,1/2]$, and $[(1-\delta')/2,(1-\delta')/2]$ and
$[(1+\delta')/2,(1+\delta')/2]$ for low, intermediate, and high
energies, respectively. These are exactly positions of the IC
peaks at both low and high energies and resonance peak at
intermediate energy determined by the dispersion of very well
defined renormalized spin excitations. Since the essential physics
is dominated by the dressed spin self-energy renormalization due
to the dressed holon bubble in the dressed holon particle-particle
channel, then in this sense the mobile dressed holon pairs (then
the electron Cooper pairs) are the key factor leading to the IC
magnetic scattering peaks at both low and high energies and
commensurate resonance peak at intermediate energy, i.e., the
mechanism of the IC magnetic scattering and commensurate resonance
in the SC-state is most likely related to the motion of the
dressed holon pairs (then the electron Cooper pairs). This is why
the position of the IC magnetic scattering peaks and commensurate
resonance peak in the SC-state can be determined in the present
study within the $t$-$J$ model under the kinetic energy driven SC
mechanism, while the dressed spin energy dependence is ascribed
purely to the self-energy effects which arise from the the dressed
holon bubble in the dressed holon particle-particle channel. Our
present result in the SC-state and the previous result in the
normal-state \cite{feng2,feng5} show that the IC magnetic
scattering at low energy appears in both SC- and normal-states,
this indicates that the IC magnetic scattering at low energy is
not associated with the SC-state, which is similar to stripe
models \cite{stock,zannen,carlson,scheidl}, where the IC magnetic
scattering at low energy is due to the formation of magnetic
domain lines \cite{stock,zannen,carlson,scheidl}. Since the
commensurate $[1/2,1/2]$ resonance at intermediate energy and IC
magnetic scattering at high energy are absent from the
normal-state \cite{feng2,feng5}, then our present result also show
that the commensurate $[1/2,1/2]$ resonance at intermediate energy
and IC magnetic scattering at high energy are closely related to
the SC-state, which is different from the stripe theory
\cite{stock,carlson,scheidl}, where the linear spin wave models
based the stripe ground state predict that the spin excitations at
high energy are nearly symmetric around $[1/2,1/2]$ position and
disperse almost linearly with energy \cite{stock,carlson,scheidl},
then the commensurate $[1/2,1/2]$ resonance may represent a
characteristic energy defined by the size of a stripe domain.
Although there are some subtle differences between our present
approach and stripe theory, both theories can give the qualitative
interpretation for all main features of the unusual spin response
of cuprate superconductors
\cite{dai,bourges0,bourges,arai,hayden,stock,tranquada}.

\section{Summary and discussions}

In summary, within the framework of the kinetic energy driven
superconductivity \cite{feng1}, we have discussed the magnetic
nature of cuprate superconductors. It is shown that the SC-state
is controlled by both SC gap function and quasiparticle coherent
weight. This quasiparticle coherent weight is closely related to
the dressed holon self-energy from the dressed spin pair bubble,
and grows linearly with the hole doping concentration in the
underdoped and optimally doped regimes, and then decreases with
doping in the overdoped regime, which leads to that the maximal SC
transition temperature $T^{(d)}_{c}$ occurs around the optimal
doping $x_{{\rm opt}}\approx 0.18$, and then decreases in both
underdoped and overdoped regimes, in qualitative agreement with
the experiments \cite{tallon}. Although the symmetry of the
SC-state is doping dependent, the SC-state has the d-wave symmetry
in a wide range of doping. Within this d-wave SC-state, we have
calculated the dynamical spin structure factor of cuprate
superconductors in terms of the collective mode in the dressed
holon particle-particle channel, and reproduce all main features
of inelastic neutron scattering experiments in the SC-state
\cite{dai,bourges0,bourges,arai,hayden,stock,tranquada}, including
the energy dependence of the IC magnetic scattering peaks at both
low and high energies and commensurate resonance peak at
intermediate energy. In particular, we have shown that the unusual
IC magnetic excitations at high energy have energies greater than
the dressed holon pairing energy (then SC Cooper pairing energy),
and are present at the SC transition temperature.

The $t$-$J$ model is characterized by a competition between the
kinetic energy ($t$) and magnetic energy ($J$). The magnetic
energy $J$ favors the magnetic order for spins, while the kinetic
energy $t$ favors delocalization of holes and tends to destroy the
magnetic order. Therefore the introduction of the additional
second neighbor hopping $t'$ in the $t$-$J$ model may be
equivalent to increase the kinetic energy, and this $t'$ term does
not change spin configuration because of the same sublattice
hopping. In this case, we \cite{feng6} have discussed the effect
of the additional second neighbor hopping $t'$ on
superconductivity within the $t$-$t'$-$J$ model, and found that
the d-wave SC pairing correlation is enhanced, while the s-wave SC
pairing correlation is heavily suppressed.

Superconductivity in cuprates emerges when charge carriers, holes
or electrons, are doped into Mott insulators
\cite{kastner,tokura}. Both hole-doped and electron-doped cuprate
superconductors have the layered structure of the square lattice
of the CuO$_{2}$ plane separated by insulating layers
\cite{kastner,tokura}. In particular, the symmetry of the SC order
parameter is common in both case \cite{tsuei,tsuei6}, manifesting
that two systems have similar underlying SC mechanism. On the
other hand, the strong electron correlation is common for both
hole-doped and electron-doped cuprates, then it is possible that
superconductivity in electron-doped cuprates is also driven by the
kinetic energy as in hole-doped case. Within the $t$-$t'$-$J$
model, we \cite{feng7} have discussed this issue, and found that
superconductivity appears around the optimal doping in
electron-doped cuprates, and the maximum achievable SC transition
temperature is lower than hole-doped cuprates due to the
electron-hole asymmetry.

\acknowledgments The author would like to thank Dr. Ying Liang,
Dr. Bin Liu, Dr. Jihong Qin, Professor Y.J. Wang, and Professor
H.H. Wen for the helpful discussions. This work was supported by
the National Natural Science Foundation of China under Grant Nos.
10125415 and 90403005.


\begin{references}

\bibitem{anderson2} P.W. Anderson, Science {\bf 235}, 1196
(1987).

\bibitem{anderson3} P.W. Anderson, Phys. Rev. Lett. {\bf 67}, 2092
(1991); Science {\bf 288}, 480 (2000); Physica C {\bf 341-348}, 9
(2000); cond-mat/0108522.

\bibitem {laughlin} R.B. Laughlin, Phys. Rev. Lett. {\bf 79},
1726 (1997); J. Low. Tem. Phys. {\bf 99}, 443 (1995).

\bibitem{kastner} See, e.g., M.A. Kastner, R.J. Birgeneau, G. Shiran,
and Y. Endoh, Rev. Mod. Phys. {\bf 70}, 897 (1998).

\bibitem{tallon} See, e.g., J.L. Tallon, J.W. Loram, J.R. Cooper,
C. Panagopoulos, and C. Bernhard, Phys. Rev. B {\bf 68}, 180501
(2003).

\bibitem{ding} H. Ding, J.R. Engelbrecht, Z. Wang, J.C. Campuzano,
S.C. Wang, H.B. Yang, R. Rogan, T. Takahashi, K. Kadowaki, and
D.G. Hinks, Phys. Rev. Lett. {\bf 87}, 227001 (2001); R.H. He,
D.L. Feng, H. Eisaki, J.-I. Shimoyama, K. Kishio, and G.D. Gu,
Phys. Rev. B {\bf 69}, 220502 (2004).

\bibitem{dai} P. Dai, H.A. Mook, R.D. Hunt, and F. Do\~gan,
Phys. Rev. B{\bf 63}, 54525 (2001); H. He, P. Bourges, Y. Sidis,
C. Ulrich, L.P. Regnault, S. Pailh\'es, N.S. Berzigiarova, N.N.
Kolesnikov, and B. Keimer, Science {\bf 295}, 1045 (2002); N.B.
Christensen, D.F. McMorrow, H.M. R\o nnow, B. Lake, S.M. Hayden,
G. Aeppli, T.G. Perring, M. Mangkorntong, N. Nohara, and H.
Tagaki, Phys. Rev. Lett. {\bf 93}, 147002 (2004).

\bibitem{yamada} K. Yamada, C.H. Lee, K. Kurahashi, J. Wada, S.
Wakimoto, S. Ueki, H. Kimura, Y. Endoh, S. Hosoya, and G. Shirane,
Phys. Rev. B {\bf 57}, 6165 (1998).

\bibitem{wakimoto} S. Wakimoto, H. Zhang, K. Yamada, I. Swainson,
H. Kim, and R.J. Birgeneau, Phys. Rev. Lett. {\bf 92}, 217004
(2004); M. Fujita, K. Yamada, H. Hiraka, P.M. Gehring, S.H. Lee,
S. Wakimoto, and G. Shirane, Phys. Rev. B {\bf 65}, 064505 (2002);
S. Wakimoto, G. Shirane, Y. Endoh, K, Hirota, S. Ueki, Y.S. Lee,
P.M. Gehring, and S.H. Lee, Phys. Rev. B {\bf 60}, R769 (1999).

\bibitem{bourges0} P. Bourges, B. Keimer, S. Pailh\'es, L.P.
Regnault, Y. Sidis, and C. Ulrich, Physica C {\bf 424}, 45 (2005);
H. He, Y. Sidis, P. Bourges, G.D. Gu, A. Ivanov, N. Koshizuka, B.
Liang, C.T. Lin, L.P. Regnault, E. Schoenherr, and B. Keimer,
Phys. Rev. Lett. {\bf 86}, 1610 (2001).

\bibitem{bourges} P. Bourges, Y. Sidis, H.F. Fong, L.P. Regnault,
J. Bossy, A. Ivanov, and B. Keimer, Science {\bf 288}, 1234
(2000).

\bibitem{arai} M. Arai, T. Nishijima, Y. Endoh, T. Egami, S.
Tajima, K. Tomimoto, Y. Shiohara, M. Takahashi, A. Garret, and
S.M. Bennington, Phys. Rev. Lett. {\bf 83}, 608 (1999).

\bibitem{hayden} S.M. Hayden, H.A. Mook, P. Dai, T.G. Perring,
and F. Do\~gan, Science {\bf 429}, 531 (2004).

\bibitem{stock} C. Stock, W.J. Buyers, R.A. Cowley, P.S. Clegg,
R. Coldea, C.D. Frost, R. Liang, D. Peets, D. Bonn, W.N. Hardy,
and R.J. Birgeneau, Phys. Rev. B{\bf 71}, 24522 (2005).

\bibitem{tranquada} J.M. Tranquada, H. Woo, T.G. Perring, H.
Goka, G.D. Gu, G. Xu, M. Fujita, and K. Yamada, Nature {\bf 429},
534 (2004); J.M. Tranquada, cond-mat/0512115.

\bibitem{hinkov} V. Hinkov, S. Pailh\'es, P. Bourges, Y. Sidis,
A. Ivanov, A. Kulakov, C.T. Lin, D.P. Chen, C. Bernhard, and B.
Keimer, Science {\bf 430}, 650 (2004).

\bibitem{feng1} Shiping Feng, Phys. Rev. B {\bf 68}, 184501
(2003).

\bibitem{feng2} Shiping Feng, Jihong Qin, and Tianxing Ma, J.
Phys. Condens. Matter {\bf 16}, 343 (2004); Shiping Feng, Tianxing
Ma, and Jihong Qin, Mod. Phys. Lett. B{\bf 17}, 361 (2003).

\bibitem{feng0} Tianxing Ma, Huaiming Guo, and Shiping Feng, Mod.
Phys. Lett. B{\bf 18}, 895 (2004).

\bibitem{eliashberg} G.M. Eliashberg, Sov. Phys. JETP {\bf 11},
696 (1960); D.J. Scalapino, J.R. Schrieffer, and J.W. Wilkins,
Phys. Rev. {\bf 148}, 263 (1966).

\bibitem{mahan} See, e.g., G.D. Mahan, {\it Many Particle
Physics}, (Plenum Press, New York, 1981), Chapter 9.

\bibitem{shen1} Z.X. Shen, D.S. Dessau, B.O. Wells, D.M. King,
W.E. Spicer, A.J. Arko, D. Marshall, L.W. Lombardo, A. Kapitulnik,
P. Dickinson, S. Doniach, J. DiCarlo, T. Loeser, and C.H. Park,
Phys. Rev. Lett. {\bf 70}, 1553 (1993); H. Ding, M.R. Norman, J.C.
Campuzano, M. Randeria, A.F. Bellman, T. Yokoya, T. Takahashi, T.
Mochiku, and K. Kadowaki, Phys. Rev. B{\bf 54}, R9678 (1996).

\bibitem{feng3} Shiping Feng and Yun Song, Phys. Rev. B {\bf 55},
642 (1997).

\bibitem{kondo} J. Kondo and K. Yamaji, Prog. Theor. Phys.
{\bf 47}, 807 (1972).

\bibitem{feng4} Shiping Feng and Zhongbing Huang, Phys. Lett. A
{\bf 232}, 293 (1997); Feng Yuan, Jihong Qin, Shiping Feng, and
W.Y. Chen, Phys. Rev. B {\bf 67}, 134505 (2003).

\bibitem{feng9} Huaiming Guo and Shiping Feng, cond-mat/0509508.

\bibitem{shen2} Z.X. Shen and D.S. Dessau, Phys. Rep. {\bf 253},
2 (1985).

\bibitem{chaudhari} P. Chaudhari and S.Y. Lin, Phys. Rev. Lett.
{\bf 72}, 1084 (1994); D.H. Wu, J. Mao, S.N. Mao, J.L. Peng, X.X.
Xi, T. Venkatesan, R.L. Greene, and S.M. Anlage, Phys. Rev. Lett.
{\bf 70}, 85 (1993); S.M. Anlage, B.W. Langley, G. Deutscher, J.
Halbritter, and M.R. Beasley, Phys. Rev. B {\bf 44}, 9764 (1991).

\bibitem{martindale} J.A. Martindale, S.E. Barrett, K.E. \'OHara,
C.P. Slichter, W.C. Lee, and D.M. Ginsberg, Phys. Rev. B {\bf 47},
9155 (1993); W.N. Hardy, D.A. Bonn, D.C. Morgan, R. Liang, and K.
Zhang, Phys. Rev. Lett. {\bf 70}, 3999 (1994); D.A. Wollman, D.J.
Van Harlingen, W.C. Lee, D.M. Ginsberg, and A.J. Leggett, Phys.
Rev. Lett. {\bf 71}, 2134 (1993).

\bibitem{tsuei} See, e.g., C.C. Tsuei and J.R. Kirtley, Rev. Mod.
Phys. {\bf 72}, 969 (2000).

\bibitem{scalettar} E. Dagotto and J. Riera, Phys. Rev. B
{\bf 46}, 12084 (1992); R.T. Scalettar, Physica C {\bf 162-164},
313 (1989); R.T. Scalettar, S.R. White, D.J. Scalapino, and R.
Sugar, Phys. Rev. B {\bf 44}, 770 (1991); M. Capone, M. Fabrizio,
C. Castellani, and T. Tosatti, Science {\bf 296}, 2364 (2002).

\bibitem{wen} H.H. Wen, H.P. Yang, S.L. Li, X.H. Zeng, A.A.
Soukiassian, W.D. Si, and X.X. Xi, Europhys. Lett. {\bf 64}, 790
(2003).

\bibitem{yeh} N.-C. Yeh, C.T. Chen, G. Hammerl, J. Mannhart, A.
Schmehl, C.W. Schneider, R.R. Schulz, S. Tajima, K. Yoshida, D.
Garrigus, and M. Strasik, Phys. Rev. Lett. {\bf 87}, 087003
(2001); G. Deutscher, Nature {\bf 397}, 410 (1999).

\bibitem{biswas} A. Biswas, P. Fournier, M.M. Qazilbash, V.N.
Smolyaninova, H. Balci, and R.L. Greene, Phys. Rev. Lett. {\bf
88}, 207004 (2002).

\bibitem{tsuei1} C.C. Tsuei, J.R. Kirtley, G. Hammerl, J.
Mannhart, H. Raffy, and Z.Z. Li, Phys. Rev. Lett. {\bf 93}, 187004
(2004).

\bibitem{uemura} Y.J. Uemura, G.M. Luke, B.J. Sternlieb, J.H.
Brewer, J.F. Carolan, W.N. Hardy, R. Kadono, J.R. Kempton, R.F.
Kiefl, S.R. Kreitzman, P. Mulhern, T.M. Riseman, D.L. Williams,
B.X. Yang, S. Uchida, H. Takagi, J. Gopalakrishnan, A.W. Sleight,
M.A. Subramanian, C.L. Chien, M.Z. Cieplak, G. Xiao, V.Y. Lee,
B.W. Statt, C.E. Stronach, W.J. Kossler, and X.H. Yu, Phys. Rev.
Lett. {\bf 62}, 2317 (1989); Y.J. Uemura, L.P. Le, G.M. Luke, B.J.
Sternlieb, W.D. Wu, J.H. Brewer, T.M. Riseman, C.L. Seaman, M.B.
Maple, M. Ishikawa, D.G. Hinks, J.D. Jorgensen, G. Saito, and H.
Yamochi, Phys. Rev. Lett. {\bf 66}, 2665 (1991).

\bibitem{dewilde} Y. DeWilde, N. Miyakawa, P. Guptasarma, M.
Iavarone, L. Ozyuzer, J.F. Zasadzinski, P. Romano, D.G. Hinks, C.
Kendziora, G.W. Crabtree, and K.E. Gray, Phys. Rev. Lett. {\bf
80}, 153 (1998).

\bibitem{feng5} Shiping Feng and Zhongbing Huang, Phys. Rev. B
{\bf 57}, 10328 (1998); Feng Yuan et al., Phys. Rev. B {\bf 64},
224505 (2001).

\bibitem{shamoto} S. Shamoto, M. Sato, J.M. Tranquada, B.J.
Sternlib, and G. Shirane, Phys. Rev. B {\bf 48}, 13817 (1993).

\bibitem{zannen} J. Zaanen and O. Gunnarsson, Phys. Rev. B
{\bf 40}, 7391 (1989); D. Poilblanc and T.M. Rice, Phys. Rev. B
{\bf 39}, 9749 (1989).

\bibitem{carlson} E.W. Carlson, D.X. Yao, and D.K. Campbell, Phys.
Rev. B {\bf 70}, 064505 (2004); C.D. Batista, G. Ortiz, and A.V.
Balatsky, Phys. Rev. B {\bf 64}, 172508 (2001).

\bibitem{scheidl} F. Kr\"uger and S. Scheidl, Phys. Rev. B
{\bf 67}, 134512 (2003).

\bibitem{feng6} Shiping Feng and Tianxing Ma, Phys. Lett. A, to
be published, cond-mat/0506114.

\bibitem{tokura} Y. Tokura, H. Takagi, and S. Uchida, Nature
{\bf 337}, 345 (1989); L. Alff, Y. Krockenberger, B. Welter, M.
Schonecke, R. Gross, D. Manske, and M. Naito, Nature {\bf 422},
698 (2003).

\bibitem{tsuei6} C.C. Tsuei and J.R. Kirtley, Phys. Rev. Lett.
{\bf 85}, 182 (2000).

\bibitem{feng7} Tianxing Ma and Shiping Feng, Phys. Lett.
A{\bf 339}, 131 (2005).

\end{references}
\end{document}